\documentclass[%
 reprint,
nofootinbib,
 amsmath,amssymb,
 aps,
 pra,
]{revtex4-1}

\usepackage{ctable}
\usepackage{multirow}
\usepackage{ulem}
\usepackage{graphicx}
\usepackage{dcolumn}
\usepackage{bm}

\usepackage{xcolor}
\usepackage{amsmath}
\usepackage[encapsulated]{CJK}

\newcommand{\ket}[1]{| #1 \rangle}

\newcommand{\add}[1]{{#1}}
\newcommand{\move}[1]{{#1}}

\begin{document}
\begin{CJK*}{UTF8}{gbsn}
\title{Quantum Optimization with a Novel Gibbs Objective Function \\and Ansatz Architecture Search}

\author{Li Li (李力)$^{1}$}
\email[email: ]{leeley@google.com}

\author{Minjie Fan (范敏杰)$^{1}$}

\author{Marc Coram$^{1}$}

\author{Patrick Riley$^{1}$}

\author{Stefan Leichenauer$^{2}$}
\affiliation{
$^1$ Google Research, Mountain View, CA 94043, USA \\
$^2$ X, The Moonshot Factory, Mountain View, CA 94043, USA}

\date{\today}

\begin{abstract}
The Quantum Approximate Optimization Algorithm (QAOA) is a standard method for combinatorial optimization with a gate-based quantum computer. The QAOA consists of a particular ansatz for the quantum circuit architecture, together with a prescription for choosing the variational parameters of the circuit. We propose modifications to both. First, we define the Gibbs objective function and show that it is superior to the energy expectation value for use as an objective function in tuning the variational parameters. Second, we describe an Ansatz Architecture Search (AAS) algorithm for searching the discrete space of quantum circuit architectures near the QAOA to find a better ansatz. Applying these modifications for a complete graph Ising model results in a $244.7\%$ median relative improvement in the probability of finding a low-energy state while using $33.3\%$ fewer two-qubit gates. For Ising models on a 2d grid we similarly find $44.4\%$ median improvement in the probability with a $20.8\%$ reduction in the number of two-qubit gates.
This opens a new research field of quantum circuit architecture design for quantum optimization algorithms.
\end{abstract}

\maketitle
\end{CJK*}

\section{Introduction}

The Quantum Approximate Optimization Algorithm (QAOA)~\cite{farhi2014quantum, farhi2014bquantum} is a general-purpose algorithm for finding a low-energy state of a given computational-basis Hamiltonian. This is a classical problem which can be combinatorially difficult, but using a quantum computer to find the solution might be more efficient than a classical method. The QAOA has performance guarantees in certain combinatorial problems~\cite{farhi2014bquantum} and quantum state transfer~\cite{niu2019optimizing}, and it has been shown that in general the output of the QAOA is not classically simulable~\cite{farhi2016quantum}. The QAOA and related algorithms offer a promising avenue for near-term applications of quantum computers~\cite{preskill2018quantum}.

\begin{figure}[htb]
\includegraphics[width=0.8\columnwidth]{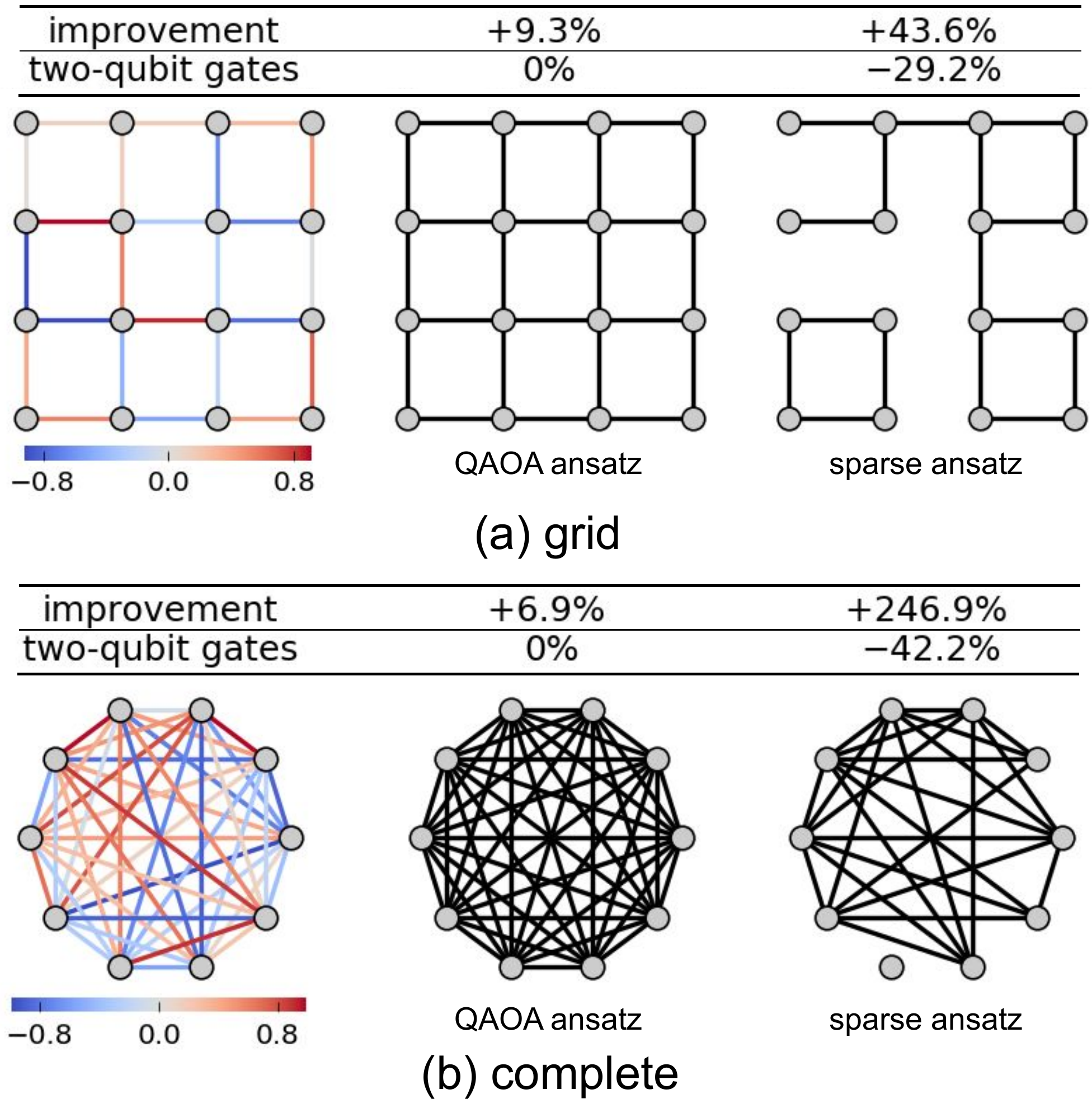}
\caption{\label{fig:plot_instances} Particular instances of random couplings (left) and the structures of the associated QAOA ansatzes (middle) and best sparse ansatzes (right) for (a) grid and (b) complete graph problems with the Gibbs objective function.
On the left, each edge in the instance graph is colored by its coupling from blue ($-1$) to red ($1$).
On the middle and right, each edge denotes the existence of two-qubit gate on the corresponding edge in the ansatz graph.
We show the relative improvement of the probability of low energy and reduction of the number of two-qubit gates compared to the usual prescription of the QAOA.
}
\end{figure}

\add{As emphasized in the original QAOA paper, } the correct way to frame the goal of quantum optimization is in the probably approximately correct framework~\cite{Valiant1984-sm}. That is, the goal is to obtain a high likelihood of finding a nearly optimal solution.
However, the standard objective function for QAOA does not reflect this goal. We introduce a new Gibbs objective function and show its superiority in the probably approximately correct sense.
In numerical experiments for grid and complete graph Ising models, using the Gibbs objective function results in $10.8\%$ and $8.6\%$ median relative improvement of the probability of finding a low-energy state, respectively.
We then proceed to try and find a superior circuit ansatz for the Gibbs objective function that is closely related to the general QAOA circuit through Ansatz Architecture Search (AAS).
Using AAS, the median relative improvement increased to $44.4\%$ and $244.7\%$ for the grid and complete graph models, together with a median reduction in the number of two-qubit gates by $20.8\%$ and $33.3\%$, respectively.
Figure~\ref{fig:plot_instances} shows two exemplary instances and the improvement of probability of low energy with ansatzes found by AAS with the Gibbs objective function.
The existence of these superior circuits opens a new field of research to design a search procedure for optimal problem-specific circuits.

\section{Ising Models}

\begin{figure}[htb]
\includegraphics[width=\columnwidth]{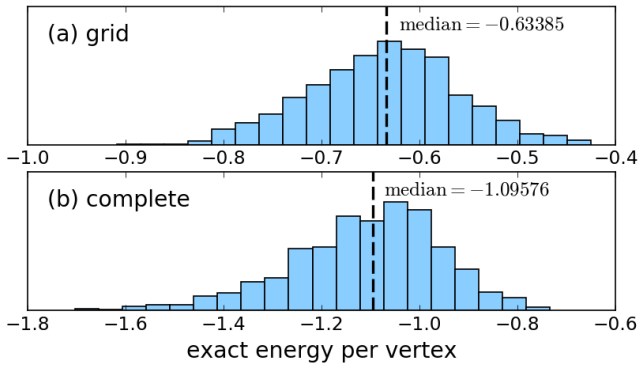}
\caption{\label{fig:exact_energy} \move{Exact ground state energies per vertex of (a) grid and (b) complete graph instances. Black dashed lines indicate the medians of the exact energies per vertex.}}
\end{figure}

A model $\mathcal{I}$ is defined on a graph $\mathcal{G}^\mathcal{I}$ with $n$ vertices $\bm{v}\in\{1,2,\dots,n\}$ and a set of undirected edges $\mathcal{E}=\{\bm{e}_{ij}\}$.
We select $4\times4$ grid and complete graph with 10 vertices to cover the extreme cases of sparse and dense graphs:
\move{
\begin{description}
    \item[Grid] A $4\times4$ square lattice. Edges only exist between nearest-neighbor vertices. This graph contains 16 vertices and $|\mathcal{E}|=24$ edges. The average degree of the vertex in this graph is 3.
    \item[Complete Graph] A complete graph with 10 vertices. Edges exist between any pair of vertices. This graph contains $|\mathcal{E}|=45$ edges. The degree of each vertex in this graph is 9.
\end{description}}
Each instance consists of a set of couplings $\bm{J}$ sampled independently from a uniform distribution $J_{ij} \sim U(-1, 1)$.
A coupling $J_{ij}$ is assigned to each undirected edge $\bm{e}_{ij}$ between vertices $i$ and $j$.
The Hamiltonian is written as a sum over edges, $E = \sum_{\bm{e}_{ij}}J_{ij}Z_i Z_j$.
We denote a problem instance as $\mathcal{I}=\mathcal{I}(\mathcal{G}^\mathcal{I},\bm{J})$. \move{In Figure \ref{fig:exact_energy} we plot histograms of the exact ground state energies per vertex for these instances.}

The QAOA specifies a particular quantum circuit architecture which depends on the Hamiltonian. The prescription is very similar to a discretized adiabatic algorithm. The quantum state produced by the QAOA at level $p$ is
\begin{equation}\label{eq:qaoa_ansatz}
\ket{\psi} = e^{i\beta_p X}e^{i\gamma_p E}\cdots e^{i\beta_1 X}e^{i\gamma_1 E}H^{\otimes n}\ket{0^n}
\end{equation}
The $2p$ parameters $\vec{\beta}$ and $\vec{\gamma}$ are variational parameters of the model. For the Ising models in this paper, only two-qubit gates are required to construct the circuit for $e^{i \gamma E}$. We will also focus on $p=1$ for simplicity.

Success in approximate optimization is measured according to the \textit{probability of finding a low energy state}. With that in mind, we evaluate the performance of our quantum circuits according to $P(E < E_0)$, where $P$ is the Born probability distribution of the output quantum state $\ket{\psi}$ and $E_0$ is the cutoff for what we consider low energy. For definiteness, in this paper we use $E_0 = 0.95 E_{\text{gs}}(\mathcal{I})$ as our definition, where $E_{\text{gs}}(\mathcal{I})$ is the exact ground state energy of the given instance $\mathcal{I}$ (which is always negative for the models we consider).

\section{Gibbs Objective Function}
\subsection{Theory}
We first address the problem of choosing the optimal values of the variational parameters.
The standard prescription of minimizing the expectation value of the energy, $\langle E \rangle$, is just a proxy for maximizing $P(E < E_0)$.
Recent work~\cite{barkoutsos2019improving} has explored using Conditional Value-at-Risk (CVaR) as the objective function.
As an alternative, we propose minimizing the \textit{Gibbs objective function}, defined as follows:
\begin{equation}\label{eq-gibbs}
f = - \log \langle e^{-\eta E} \rangle.
\end{equation}
Here $\eta > 0$ is a hyperparameter based on the general properties of the class of problems. The function $f$ is very similar to the Gibbs free energy from statistical mechanics, which is the origin of the name.

The reason why $\langle e^{-\eta E} \rangle$ might be preferred over $\langle E \rangle$ is easily understood intuitively. The exponential profile rewards us for increasing the probability of low energy, and de-emphasizes the shape of the probability distribution at higher energies.
Note that the Gibbs objective function is just as easy to measure as the energy expectation value itself when the energy is diagonal in the computational basis: we just perform a different computation with our measurement samples.

The Gibbs objective function is essentially the cumulant generating function of the energy~\cite{McCullagh:2009}. The Taylor expansion reads $f(\eta) = \mu_E\eta - \sigma_E^2\eta^2/2 + \kappa_3 \eta^3/6 + \cdots.$ For small $\eta$, then, minimizing the Gibbs objective function is equivalent to minimizing $\mu_E = \langle E \rangle$. As $\eta$ increases, the higher-order cumulants become more important.

\move{To better understand the Gibbs objective function, we can try to estimate the best value of the hyperparameter $\eta$. For any $\eta >0$ the probability of low energy is bounded from above as follows:
\begin{align}
P(E < E_0)  &= \langle 1_{E < E_0} \rangle \nonumber \\ 
&\leq \langle 1_{E <E_0} e^{-\eta(E-E_0)}\rangle \nonumber \\
&\leq \langle e^{-\eta(E-E_0)}\rangle.
\end{align}
Choosing $\eta$ to minimize the right-hand side gives the strongest inequality out of this one-parameter family. That value of $\eta$ is the one which satisfies the equation
\begin{equation}\label{eq-optimaleta}
E_0 =\frac{\langle E e^{-\eta E}\rangle}{\langle e^{-\eta E}\rangle}.
\end{equation}
Now, $\eta$ is meant to be a fixed hyperparameter that is maintained throughout parameter optimization, whereas the $\eta$ satisfying this equation depends functionally on the probability distribution itself. Our prescription for estimating $\eta$ is to find an approximate solution to this equation, valid for a large class of probability distributions that we may encounter during parameter optimization.}

\move{If $E_0$ is meant to be close to $E_\text{gs}$, then it's clear that the interesting limit of Eq.~\eqref{eq-optimaleta} is the large-$\eta$ limit.\footnote{We are assuming that $\eta$ is large compared to the inverse of the energy scale of the Hamiltonian, but not large compared to the gap. In other words, even when $\eta$ is large there should still be many states between $E_\text{gs}$ and $E_\text{gs} + \eta^{-1}$.} The first correction at large-$\eta$ to the RHS is equal to $\eta^{-1}$:
\[
\frac{\langle E e^{-\eta E}\rangle}{\langle e^{-\eta E}\rangle}
\approx E_\text{gs} + \eta^{-1}.
\]
Combined with Eq.~\eqref{eq-optimaleta}, this suggests that we should set $\eta = (E_0 - E_\text{gs})^{-1}$. We may only be able to estimate values for $E_0$ and $E_\text{gs}$ based on the specification of our problem, but in practice these estimates are good enough. For the problems we consider, $E_0 = 0.95 E_\text{gs}$ and $E_\text{gs} \approx -1$ gives $\eta \approx 20$ as an estimate, which we use for the majority of our numerical experiments below.}

\move{Note that in the large-$\eta$/small-$(E_0 - E_\text{gs})$ regime one can make much stronger statements about the relationship between $\langle e^{-\eta E}\rangle$ and $P(E < E_0)$. We will sketch some of them here. In taking the large-$\eta$ limit above, we effectively approximated the probability density function for the energy, $p(E)$, by its constant term $p(E_\text{gs})$ near the ground-state energy. If $p(E)$ is treated as a constant, then $P(E < E_0)$ and $\langle e^{-\eta (E - E_\text{gs})}\rangle$ are actually \textit{equal} when $\eta = (E_0 - E_\text{gs})^{-1}$. More generally, if $p(E)$ is well-approximated by a finite-degree polynomial in $E-E_\text{gs}$ with bounded coefficients, then we have the slightly weaker condition $P(E < E_0) \sim\langle e^{-\eta (E - E_\text{gs})}\rangle$, meaning that either quantity is bounded from above and below by constant multiples of the other. This further motivates the use of the Gibbs objective function.}

\subsection{Numerical Experiments}

To evaluate the performance of the Gibbs objective function, as well as the ansatz search described later, we analyze 1000 instances each of the grid and complete graph Ising models. For each instance we optimize the variational parameters $\beta$ and $\gamma$ using the Nelder-Mead algorithm~\cite{nelder1965simplex} to minimize either the expectation value of the energy or the Gibbs objective function. The underlying circuit ansatz is either the QAOA or an optimized sparse ansatz as described in the next section. In all cases we evaluate the algorithm performance according to the probability of finding a low energy state, $P(E < 0.95E_\text{gs}(\mathcal{I}))$. \add{As discussed above, the quantum optimization algorithm is not designed to find the exact ground state, and we chose the metric as $5\%$ around the ground state energy so the task of finding low energy is hard but not impossible. The exact value chosen here is not important.}

\begin{figure}[htb]
\includegraphics[width=0.9\columnwidth]{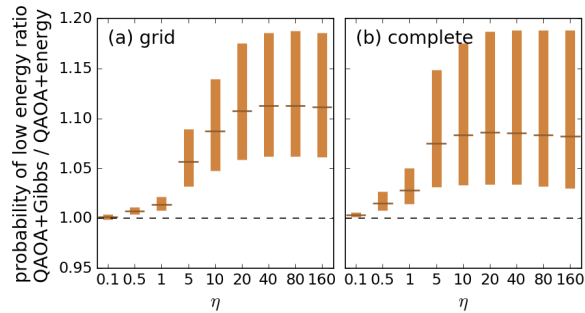}
\caption{\label{fig:different_eta} \move{Comparison of Gibbs objective function with different $\eta$ to the energy expectation objective function on QAOA ansatz. 
For every instance given $\eta$, we measure the probability of low energy of QAOA+Gibbs divided by QAOA+energy.
The bars show the range from $5\%$ to $95\%$ and the horizontal segments are median.
For small values of $\eta$, the Gibbs objective function is equivalent to the energy expectation value for purposes of optimization, while for large values of $\eta$ it is equivalent to maximizing the probability of finding the ground state.}}
\end{figure}

\move{In Figure~\ref{fig:different_eta} we show the effect of changing the hyperparamter $\eta$ in the Gibbs objective function using the QAOA circuit ansatz.}
\add{When $\eta$ is small, the Gibbs objective function is equivalent to the energy expectation value as an objective function. Therefore the Gibbs objective function cannot perform worse than the energy expectation value when $\eta$ is properly tuned.} Numerically, we find that the probability of low energy increases monotonically with $\eta$ before plateauing at large values of $\eta$. Our estimated value $\eta = 20$ falls within the convergence range for the problems we consider, so we use that value throughout. Finally, we also observe that in the extreme-$\eta$ regime, e.g., $10^5$, the parameter optimization does not converge due to the fact that the objective function is approximately zero except when the exact ground state is sampled. This is an obstacle to efficient optimization at those extreme values.

\begin{figure}[htb]
\includegraphics[width=0.9\columnwidth]{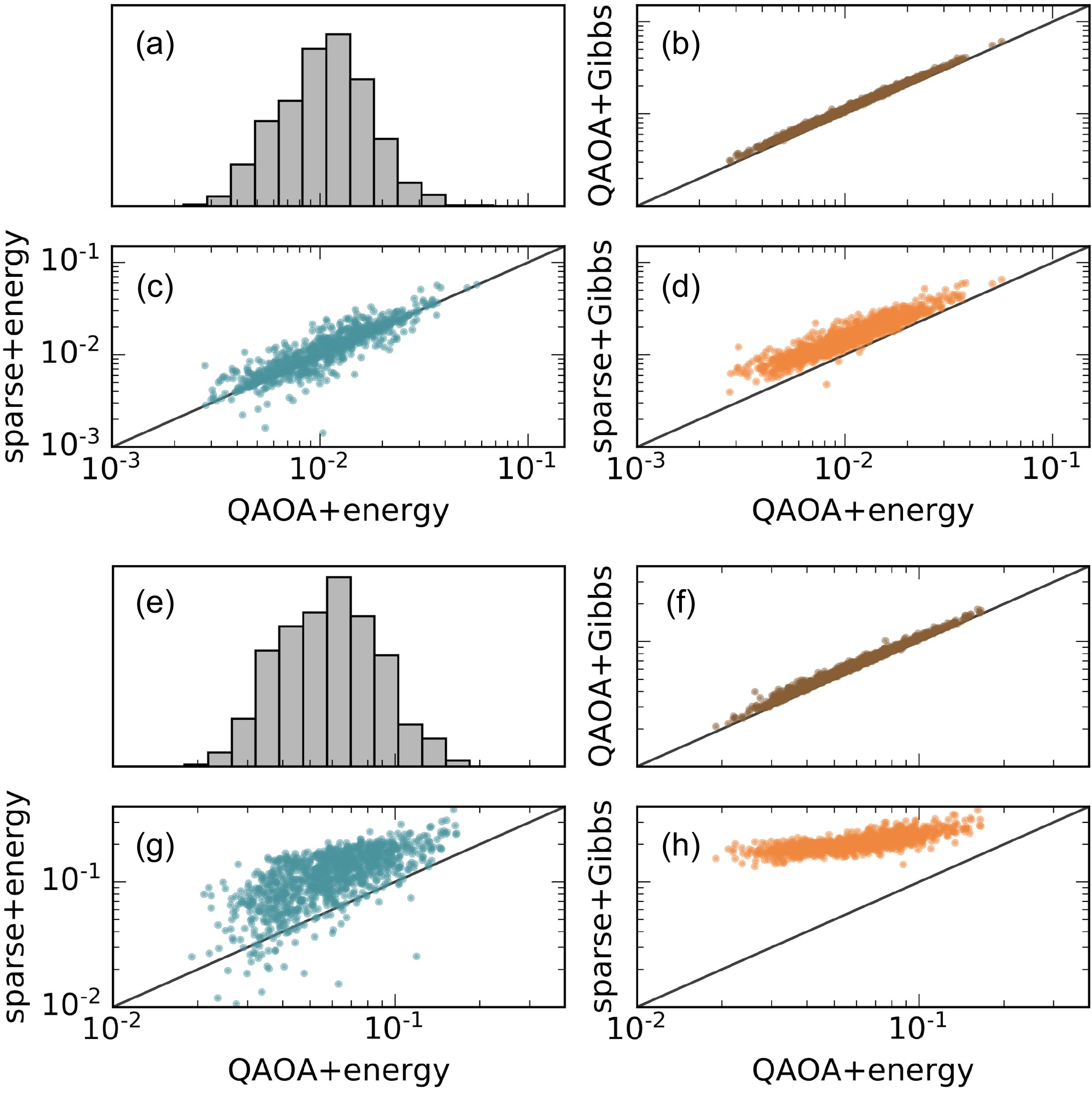}
\caption{\label{fig:compare_obj}Comparison of the objective functions and ansatzes on 1000 grid (a)-(d) and complete (e)-(h) graph instances.
The histograms show the distributions of probability of low energy for QAOA$+$energy.
The scatter plots compare the probability of low energy for \{ansatz\}$+$\{objective\} pairs against the QAOA$+$energy baseline.}
\end{figure}

Figure~\ref{fig:compare_obj} displays the probability of finding a low energy state for each quantum optimization algorithm, denoted as $\{\text{ansatz type}\}+\{\text{objective}\}$. QAOA$+$energy is the original QAOA prescription and provides the baseline for comparison. The sparse ansatz is the subject of the next section.
As shown in the scatter plots of QAOA$+$Gibbs vs. QAOA$+$energy, using the Gibbs objective function improves the solution. More significant improvement can be achieved using a sparse ansatz in addition to the Gibbs objective function, especially for complete graph instances.

\section{Optimizing the Ansatz}

Next we discuss alternatives to the QAOA circuit ansatz. For the Ising Hamiltonians, the operator $e^{i\gamma E}$ of Eq.~\eqref{eq:qaoa_ansatz} involves a two-qubit operator for each edge in the instance graph $\mathcal{G}^\mathcal{I}$. We denote by $\mathcal{G}^\mathcal{A}$ the \textit{ansatz graph}, which is obtained from $\mathcal{G}^\mathcal{I}$ by removing some edges. The associated circuit ansatz $\mathcal{A}$ is obtained by removing from $e^{i\gamma E}$ those two-qubit operators corresponding to the edges which were removed from $\mathcal{G}^\mathcal{I}$. The rest of the quantum circuit remains the same as in the QAOA. This is clearly not the most general possible prescription for $\mathcal{G}^\mathcal{A}$, but makes use of the intuition that the QAOA ansatz $\mathcal{G}^\mathcal{A} = \mathcal{G}^\mathcal{I}$ is a good starting point for the architecture search.

In total, an ansatz $\mathcal{A}(\mathcal{G}^\mathcal{A},\beta,\gamma)$ is determined by its graph architecture $\mathcal{G}^\mathcal{A}$ and continuous parameters $\beta, \gamma$. The optimal ansatz graph and variational parameters for a given instance are denoted by $\hat{\mathcal{G}}^\mathcal{A},\hat{\beta},\hat{\gamma}$, and they are the ones that minimize the objective function:
\begin{equation}\label{eq:optimal_ansatz_general}
\hat{\mathcal{G}}^\mathcal{A},\hat{\beta},\hat{\gamma} = \operatorname*{arg\,min}_{\mathcal{G}^\mathcal{A},\beta,\gamma} f(\mathcal{A}(\mathcal{G}^\mathcal{A},\beta,\gamma), \mathcal{I}).
\end{equation}
For each $\mathcal{G}^\mathcal{A}$, $\mathcal{A}(\mathcal{G}^\mathcal{A},\beta,\gamma)$ represents a family of ansatzes differing by $\beta,\gamma$. We can optimize Eq.~\eqref{eq:optimal_ansatz_general} in a nested manner,
\begin{align}
\hat{\mathcal{G}}^\mathcal{A} &= \operatorname*{arg\,min}_{\mathcal{G}^\mathcal{A}} f(\mathcal{A}(\mathcal{G}^\mathcal{A},\hat{\beta},\hat{\gamma}), \mathcal{I})\label{eq:optimal_ansatz_nested:outer} \\
\text{with  }  \hat{\beta},\hat{\gamma} &= \operatorname*{arg\,min}_{\beta,\gamma} f(\mathcal{A}(\mathcal{G}^\mathcal{A},\beta,\gamma), \mathcal{I}).\label{eq:optimal_ansatz_nested:inner}
\end{align}
The outer step (Eq.~\eqref{eq:optimal_ansatz_nested:outer}) searches the space of the architectures $\{\mathcal{G}^\mathcal{A}\}$. For a fixed architecture $\mathcal{G}^\mathcal{A}$, the inner step (Eq.~\eqref{eq:optimal_ansatz_nested:inner}) returns the optimal ansatz $\mathcal{A}(\mathcal{G}^\mathcal{A},\hat{\beta},\hat{\gamma})$ in the family of $\mathcal{A}(\mathcal{G}^\mathcal{A},\beta,\gamma)$. We should understand that $\hat{\beta}$ and $\hat{\gamma}$ are implicit functions of the ansatz graph $\mathcal{G}^\mathcal{A}$ through Eq.~\eqref{eq:optimal_ansatz_nested:inner}.
We denote the outer step as AAS and the inner step as parameter optimization.

\subsection{Ansatz Architecture Search}

A good design of the search space is essential in discrete structure optimization problems, e.g., neural architecture search~\cite{zoph2016neural,xie2019exploring,li2019random}, molecule optimization~\cite{zhou2019optimization}, composite design~\cite{gu2018bioinspired} and symbolic regression~\cite{schmidt2009distilling,Udrescu2019-cz}.
Since the QAOA is a well-recognized ansatz for combinatorial problems, we have designed the search space for $\mathcal{G}^\mathcal{A}$ based on gradual modifications of the QAOA ansatz. The QAOA prescription is to take $\mathcal{G}^\mathcal{A} = \mathcal{G}^\mathcal{I}$, and our search through architectures is a search through graphs obtained by removing edges from $\mathcal{G}^\mathcal{I}$.

Denote by $\mathcal{G}_k$ a graph containing $k$ edges. If $m$ is the number of edges in $\mathcal{G}^\mathcal{I}$, then there is only one $\mathcal{G}_m$ in our search space, namely $\mathcal{G}^\mathcal{I}$ itself. Thus we say $|\{\mathcal{G}_{m}\}|=1$. If we remove up to $n$ edges from the graph, then the total search space is
\begin{equation}
\bigcup_{l=0}^n\{\mathcal{G}_{m-l}\}=
\{\mathcal{G}_{m}\} \cup \{\mathcal{G}_{m-1}\} \cup \ldots \cup \{\mathcal{G}_{m-n}\}.
\end{equation}
The size of this space is $\sum_{l=0}^{n}\binom{m}{l}$. A brute-force enumerative search is impractical since the size grows quickly as $n$ increases.
\add{For example, considering a complete graph with 10 vertices, $\sum_{l=0}^{5}\binom{45}{l}\sim1\times10^{6}$ and $\sum_{l=0}^{15}\binom{45}{l}\sim6\times10^{11}$.}
We propose greedy search as an affordable strategy for AAS.
Given an instance $\mathcal{I}$, the search starts with $\mathcal{G}^\mathcal{A} = \mathcal{G}_{m}$ at level $0$. 
Then level by level, ansatzes are expanded by removing one two-qubit gate from the best ansatz of previous level, scored, and the best of them is selected as the output of this level.
The output architectures at level $l$ have $l$ two-qubit gates (i.e., edges of the graph) removed.
The total number of architectures visited in the greedy search is $\mathcal{N}\leq \sum_{l=0}^{n} (m-l)=(n+1)(m-n/2)$.

\subsection{Scoring an Ansatz}

\subsubsection{Nelder-Mead}

The score for each ansatz $\widetilde{\mathcal{G}}_{m-l}$ in AAS is obtained by specifying parameters $\beta^*$ and $\gamma^*$ and computing $f(\mathcal{A}(\widetilde{\mathcal{G}}_{m-l},\beta^*,\gamma^*), \mathcal{I})$.
One prescription is to let $\beta^*$ and $\gamma^*$ simply be the optimal values of $\beta$ and $\gamma$ minimizing the objective function,
$\beta^*,\gamma^* = \operatorname*{arg\,min}_{\beta,\gamma} f(\mathcal{A}(\widetilde{\mathcal{G}}_{m-l},\beta,\gamma), \mathcal{I})$.
We use the Nelder-Mead algorithm~\cite{nelder1965simplex} to perform this minimization.
In other words, we take $\beta^*$ and $\gamma^*$ to be close approximation to the optimal values $\hat{\beta}$ and $\hat{\gamma}$ for the given ansatz graph.
Nelder-Mead is a blackbox optimization algorithm popular in the quantum variational circuit literature~\cite{peruzzo2014variational,verdon2018universal}.
Using this algorithm requires running quantum circuit simulations at each iteration and reporting the objective function value to the optimizer until convergence. Thus it is extremely expensive in terms of calls to the (simulated) quantum computer. Since we want to limit the number of such calls, we are motivated to consider other strategies for finding $\beta^*$, $\gamma^*$.

\subsubsection{Estimated $\beta,\gamma$}

\move{Rather than use an optimization algorithm like Nelder-Mead to minimize the objective function and thereby obtain $\beta^*$ and $\gamma^*$, we can use analytical estimates to approximate the parameters instead. This saves the computation time required to evaluate the quantum circuits for parameter optimization during scoring. 
The $\beta^*$ and $\gamma^*$ we find will not necessarily be close to the optimal values $\hat{\beta}$ and $\hat{\gamma}$, but the idea is that this may not be important for the purposes of scoring. We may still wish to use Nelder-Mead for evaluation of the final ansatz at the conclusion of the AAS.}

\move{The estimates for $\beta^*$ and $\gamma^*$ we use in this section come from making several simplifying assumptions that are not necessarily valid. The first assumption is that it is reasonable to use $\beta^*$ and $\gamma^*$ values obtained by minimizing the energy expectation value instead of the Gibbs objective function. Focusing the grid Ising model, we can write down an exact formula for $\langle E\rangle$ in a $p=1$ QAOA as follows:
\begin{equation}\label{eq:energy}
\langle E \rangle  = \sin4\beta \sum_{\mathbf{e}_{ij}}\left[\prod_k \cos \left(2\gamma  J_{ k j }\right) \right] J_{ij}\tan \left(2\gamma  J_{ij}\right).
\end{equation}
To use this formula for an ansatz graph $\mathcal{G}^\mathcal{A}$ other than $\mathcal{G}^\mathcal{I}$ one just sets to $0$ the $J_{ij}$ associated to the missing edges.}

\move{Eq.~\eqref{eq:energy} determines $\beta^* = \pi/8$. To find a formula for $\gamma^*$ we make another simplifying assumption, namely that $\gamma^* \ll 1$.\footnote{Rather than finding an explicit formula, one could also choose to minimize Eq.~\eqref{eq:energy} numerically to find $\gamma^*$. This does not affect the results.} Expanding Eq.~\eqref{eq:energy} to third order in $\gamma$ and minimizing the resulting cubic polynomial gives
\begin{equation}\label{eq:estimated_gamma}
\gamma^*= -\sqrt{\frac{\sum_{ij} J_{i j}^2}{6\left(\sum_{ijk,j\neq k}  J_{ k i}^2 J_{ i j }^2 + \frac{1}{3}\sum_{ij} J_{ i j }^4\right)}}.
\end{equation}
All of this was in the context of grid instances, and in particular in deriving Eq.~\eqref{eq:energy} we made use of the fact that two neighboring vertices in the graph do not have any neighbors in common. Nevertheless, as our final simplifying assumption we will insist on using Eq.~\eqref{eq:estimated_gamma} for the complete graph as well. We will see from the numerical experiments that these simplifying assumptions are good enough for scoring.}

\subsubsection{Fixed $\beta,\gamma$}

The estimated $\beta^*$ above is already instance-independent, and the estimated $\gamma^*$ performed well despite the approximations involved not being fully justified. This suggests that the precise value of $\gamma^*$ used in scoring is not crucially important. This is similar to the observation in Ref.~\cite{brandao2018fixed} that the behavior of the QAOA tends to concentrate across instances. This motivates the third scoring prescription, the ``fixed-parameter'' prescription.

\begin{figure}[htb]
\includegraphics[width=\columnwidth]{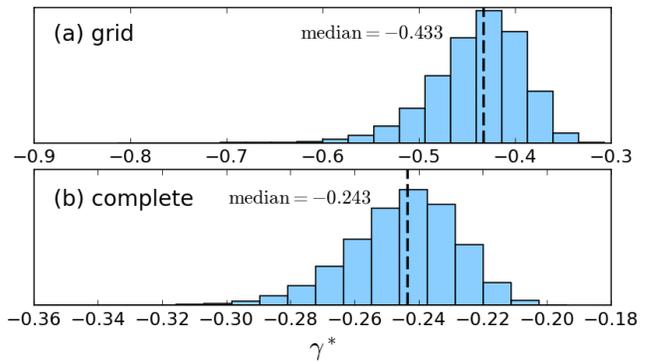}
\caption{\label{fig:fixed_parameters}
\move{Histogram of $\gamma^*$ as determined by Eq.~\eqref{eq:estimated_gamma} for $10^5$ independently drawn sets of couplings $\mathbf{J}$. Black dashed lines are the medians of the distributions.}} 
\end{figure}

\move{We generated the distribution of estimated $\gamma^*$ values according to the formula Eq.~\eqref{eq:estimated_gamma} for both the grid and complete graph models by looking at $10^5$ choices of couplings drawn independently from the uniform distribution, $\bm{J}\sim U(-1, 1)$, for each model with $\mathcal{G}^\mathcal{A} = \mathcal{G}^\mathcal{I}$. The associated histograms are shown in Figure~\ref{fig:fixed_parameters}. The ``fixed-parameter" prescription for $\gamma^*$ is defined by using the medians of these distributions for all instances of the associated model. These values are listed in Figure~\ref{fig:fixed_parameters}.}

\subsection{Numerical Experiments}

We apply the AAS procedure with the three prescriptions to choose the scoring parameters $\beta^*$ and $\gamma^*$ on grid and complete graph Ising models. 
The performance of an ansatz $\mathcal{A}$ produced by AAS is measured by the scaled probability of low energy,
\begin{equation}
\widetilde{P}(\mathcal{A},\mathcal{I})=P_{\mathcal{A}}(E < 0.95E_\text{gs})/P_\mathcal{I}(E < 0.95E_\text{gs}).
\end{equation}
The probability in the numerator is the one associated to the ansatz graph $\mathcal{G}^\mathcal{A}$ with parameters equal to their optimal values $\hat{\beta}$, $\hat{\gamma}$ obtained by minimizing the Gibbs objective function for that ansatz. The probability in the denominator is similar, except using the instance graph $\mathcal{G}^\mathcal{I}$ as the ansatz. In other words, the prescription for computing the denominator probability is similar to the standard QAOA, except the parameters are optimized using the Gibbs objective function rather than the energy expectation value. We chose to use the Gibbs objective function for both numerator and denominator in order to isolate the effects of the AAS. The optimization is done using Nelder-Mead for both.

\begin{figure}[htb]
\includegraphics[width=0.9\columnwidth]{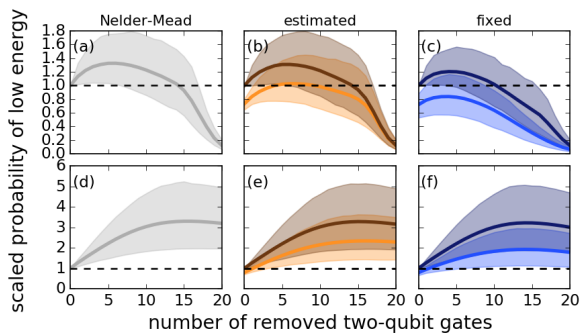}
\caption{\label{fig:two_qubits_gates_removed}
Comparison of different prescriptions for the scoring function of AAS.
The solid curves are the scaled probability of low energy of the best ansatz found through greedy search at each of the first 20 levels. The shadows show the range from 5\% to 95\%.
(a)-(c) are results from 1000 grid instances and (d)-(f) are results from 1000 complete graph instances. The scoring prescriptions are
(a)(d) Nelder-Mead, (b)(e) estimated parameters, and (c)(f) fixed parameters. The dark orange and blue curves are the scaled probability of the best ansatz graph after using Nelder-Mead to conduct a final optimization of the parameters after AAS, while the light orange (b)(e) and light blue (c)(f) curves represent the scaled probability obtained for the best ansatz graph without the final parameter optimization step.}
\end{figure}

Figure~\ref{fig:two_qubits_gates_removed} shows the scaled probabilities of low energy of the optimal ansatz at each level for both grid and complete graph instances. Each column corresponds to a different prescription for the scoring function of AAS.

We first discuss the results of grid instances.
In (a), the scoring is done using parameters that are optimized by Nelder-Mead.
The scaled probabilities of low energy increase as more two-qubit gates are removed. But they start to decrease when more than 5 two-qubit gates are removed. In (b) and (c) the scoring was performed according to the estimated parameter prescription and fixed parameter prescription, respectively. The dark curves in each case represent the performance of the final circuit found by AAS using the optimal $\hat{\beta}$, $\hat{\gamma}$ obtained by minimizing the Gibbs objective function. We see that there is not a strong dependence on the scoring prescription, though the ``fixed'' procedure is slightly worse.
However, the light curves in (b) and (c) represent the performance of those same output circuits if, rather than using $\hat{\beta}$ and $\hat{\gamma}$, we use the $\beta^*$ and $\gamma^*$ values used in the scoring step of AAS. Then we see that there is a significant decrease in performance, especially for the fixed method in (c). 
The lesson here is that for scoring in AAS, which only cares about relative performance for ranking, the circuit parameter values are less important.
In fact, good relative performance from these two prescriptions suggests that it is possible to construct inexpensive heuristic functions for scoring without calls to the quantum computer. We explore this further in Appendix~\ref{sec:challenge}.
On the other hand, it is crucial to get the parameters right when considering absolute performance.

The trend for complete graphs in (d)-(f) is very similar. The main qualitative change is that the performance does not drop off as steeply as a function of the number of removed gates. This is easily understood from the fact that the complete graphs have far more edges than the grid (45 vs 24). We also see that the ``fixed" procedure is closer in performance to the others for complete graphs.

\section{Conclusion}
We have proposed using the Gibbs objective function and AAS as two improvements to the QAOA. There are several potential follow-ups and opportunities for further developments:

The Gibbs objective function may be useful more broadly for quantum optimization problems, such as variational approaches to molecular ground states~\cite{peruzzo2014variational, mcclean2016theory}. In those cases, where the energy is not diagonal in the computational basis, it will be more challenging to evaluate $\langle \exp{(-\eta E)}\rangle$ by sampling, but may still be worthwhile.

Even within combinatorial optimization, AAS is costly because each quantum circuit must be simulated (or run on a real quantum computer) during the scoring step. Performance improvements could be offset by this extra cost. That is the motivation for the alternative heuristic methods we explore in Appendix~\ref{sec:challenge}, and it remains an open problem to find an effective heuristic. Our estimated parameter and fixed parameter prescriptions for scoring show that it is possible to capture relative performance without reproducing the absolute performance. This leaves open the possibility that a good heuristic scoring function exists. 

In this paper, we computed probabilities and expectation values directly from the wavefunction. On a real quantum computer this is impossible. Instead, one estimates expectation values based on a finite number of samples. The number of samples is another hyperparameter, and it directly affects the cost of running the algorithm on a quantum computer. An open question is whether the scoring in AAS can work with a very small number of samples, mitigating the cost. Finally, one may want to include other effects in the scoring, e.g., the fidelity of the two-qubit gates in the circuit, and search for the Pareto optimal~\cite{deb2014multi} for multi-objective optimization.\footnote{We thank Edward Farhi for discussion of this point.}

\begin{acknowledgments}
The authors thank Steven Kearnes for code review and discussions, Zan Armstrong, Nathan Neibauer for suggestions and help in data visualization, and Dave Bacon, Edward Farhi, Murphy Yuezhen Niu, Thomas F\"{o}sel, John Platt for their review and comments.
X, formerly known as Google[x], is part of the Alphabet family of companies, which includes Google, Verily, Waymo, and others (www.x.company). Quantum simulation and AAS in this paper were implemented using Cirq~\cite{cirq} and Apache Beam~\cite{apache_beam}.
\end{acknowledgments}

\appendix

\section{Greedy Search}\label{sec:greedy}

The following three steps are performed at level $l$ in the search:
\begin{description}
    \item[Expansion] Generate all the unique $\{\widetilde{\mathcal{G}}_{m-l}\}$ by removing one two-qubit gate from the output of the previous level.
    \item[Scoring] Evaluate a scoring function $\mathcal{S}$ on each of the architectures $\{\widetilde{\mathcal{G}}_{m-l}\}$ generated by the previous step. 
    Ideally, the scoring function would exactly match the final target function. However, that can be expensive to compute so we will examine alternative scoring functions below. In particular, we will consider different methods for specifying variational parameters $\beta^*, \gamma^*$ for each circuit, and then evaluating the Gibbs objective function by simulation using those parameters:
    \begin{equation}\label{eq:structure_score}
    \{\mathcal{S}(\widetilde{\mathcal{G}}_{m-l},\mathcal{I})\}=\{f(\mathcal{A}(\widetilde{\mathcal{G}}_{m-l},\beta^*,\gamma^*), \mathcal{I})\}.
    \end{equation}
    \item[Selection] Select the architecture with the best score as the output of this level.
\end{description}

\section{Initial values in Nelder-Mead}
The initial values of $\beta$ and $\gamma$ are sampled independently from the uniform distribution $U(0, 0.1)$.

\section{Beam Search}

We introduce beam search, a generalized search algorithm of greedy search.
It has been used in combinatorial optimization~\cite{li2018combinatorial}, program synthesis~\cite{bunel2018leveraging} and machine translation~\cite{sutskever2014sequence}.
Beam search differs from greedy search in the selection step:
\begin{description}
    \item[Selection] Select $w$ architectures with the best scores, where the integer $w$ is called the \textit{beam width}. These best-performing architectures are the output of this level. At early stages in the beam search we may have fewer than $w$ candidates available, in which case all candidates are returned.
\end{description}

At the $l$-th level, $|\{\widetilde{\mathcal{G}}_{m-l}\}| \leq w\times (m-l)$. The total number of architectures visited in the beam search is $\mathcal{N}\leq w\sum_{l=0}^{n} (m-l)=w(n+1)(m-n/2)$.
As special cases, we recover enumerative search as $w\to\infty$ and greedy search as $w=1$.

\begin{figure}[htb]
\includegraphics[width=\columnwidth]{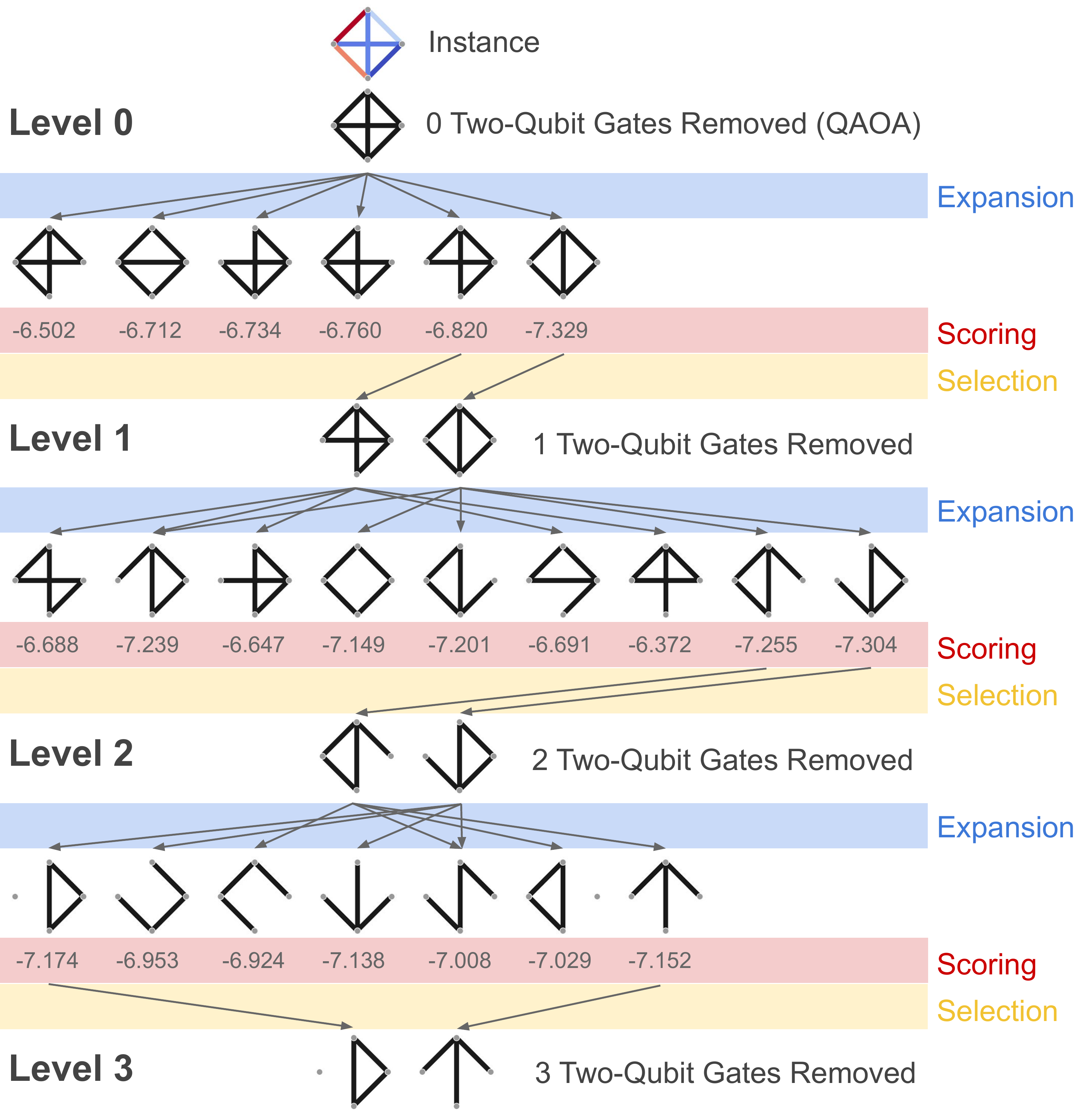}
\caption{\label{fig:ansatz_beam_search} Illustration of Ansatz Architecture Search (AAS) by removing up to 3 two-qubit gates with beam width $w=2$.}
\end{figure}

Figure~\ref{fig:ansatz_beam_search} illustrates the procedure of AAS for a complete graph with 4 vertices.

From numerical experiments, we found that there was not much improvement in performance from increasing the beam width $w\geq 1$.

\section{Challenge: Search without Quantum Simulation}\label{sec:challenge}
We demonstrated that with AAS and parameter optimization, a circuit ansatz that significantly improves the probability of low energy can be found. 
However, all of the methods in Figure~\ref{fig:two_qubits_gates_removed} made use of quantum circuit simulation at each stage in the search. While we were able to show the method with the most quantum circuit simulation (Nelder-Mead) does not improve significantly on cheaper scoring methods, all the methods require some quantum circuit simulation at each level.
In this section, we investigate some heuristic functions for replacement of quantum simulation for the purpose of the scoring step of AAS. Our results are mixed, and fully solving this problem remains an open challenge for the community.

\paragraph{Random}
For each ansatz in the scoring step, we assign a random number to replace $f(\mathcal{A},\mathcal{I})$ in Eq.~\eqref{eq:structure_score}. This baseline does not use any information from the ansatz architecture and problem instance, and amounts to removing edges from the graph randomly during AAS.

\paragraph{Energy Approximation}

Our second heuristic uses the estimated energy expectation value as the scoring function. That is, we plug $\beta^* = \pi/8$ and $\gamma^*$ as given by Eq.~\eqref{eq:estimated_gamma} into Eq.~\eqref{eq:energy} and use that as the score.\footnote{Really, we first expanded Eq.~\eqref{eq:energy} to third order in $\gamma$ before plugging in $\gamma^*$. This is for consistency, but does not make a large difference. We also experimented with numerically minimizing Eq.~\eqref{eq:energy} rather than using any estimates, and this, too, does not make much difference.}

\paragraph{Neural Network}
We use a neural network to approximate $f(\mathcal{A},\mathcal{I})$ in Eq.~\eqref{eq:structure_score}. It contains two dense layers with 128 hidden units and ReLU activation functions. The instance is represented by 2rd and 4th powers of couplings on edges. The ansatz graph $\mathcal{G}^\mathcal{A}$ is represented by booleans indicating whether a two-qubit gate is placed on an edge of the instance. We concatenate these features as input of the network.
We take all the ansatzes generated by AAS and Nelder-Mead and split them randomly by their instances into a training set and test set. For grid instances, the training set contains 800 instances with 232,800 ansatzes. For complete graph instances, the training set contains 800 instances with 568,800 ansatzes. Both test sets contain 200 instances not seen in the training set. To fix normalization we use the scaled objective function value $f(\mathcal{A},\mathcal{I})/f(\text{QAOA},\mathcal{I})$ as the label.

\begin{figure}[htb]
\includegraphics[width=\columnwidth]{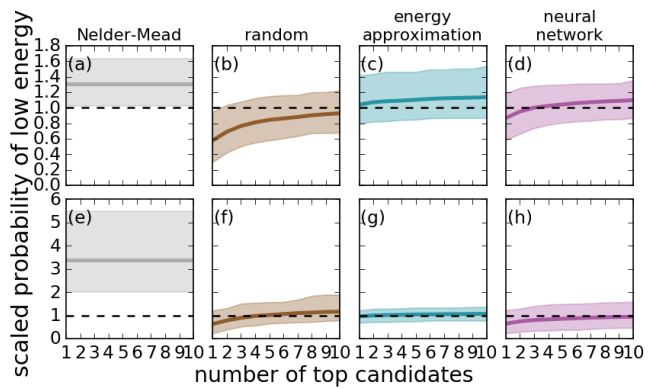}
\caption{\label{fig:baseline_with_top_candidates}
Comparison of different heuristics for the scoring function of AAS.
(a)-(d) search sparse ansatzes by removing exactly 5 two-qubit gates on 200 grid instances, and (e)-(h) search sparse ansatzes by removing exactly 15 two-qubit gates on 200 complete graph instances. 
(a)(e) use the Nelder-Mead scoring function at each level in greedy search, and serves as the baseline for measuring performance. The remaining prescriptions, explained in detail in Section~\ref{sec:challenge}, are (b)(f) random, (c)(g) energy approximation, and (d)(h) neural networks. For each of these three prescriptions we use beam width $w=100$. In all cases, Nelder-Mead is used at the end to optimize the parameters of the top candidates from the AAS (the number of which varies along the horizontal axis), and the candidate with the lowest Gibbs objective function value is selected for the plot. The solid lines are the mean performance across instances, the dashed lines are at 1, and the shadows show the range from 5\% to 95\%.
}
\end{figure}

Although a training set is not required for the random and energy approximation heuristics, for a fair comparison we restrict each heuristic to the same 200 test instances for each instance type in Figure~\ref{fig:baseline_with_top_candidates}.
We find optimal sparse ansatzes by removing 5 two-qubit gates for grid instances and 15 two-qubit gates for complete graph instances.
(a)(e) show the results of AAS and Nelder-Mead. They are the best ansatzes we can find for the test instances.
Since random, energy approximation and neural network are inexpensive to compute compared to quantum simulation, we given these an advantage in the search by setting the beam width $w$ to 100. At the end of AAS, we sort the ansatzes from the last level by their objective function values. Then we run quantum simulations for the top candidates and report the best scaled probability of low energy. As more candidates are taken into consideration, the performance of all three heuristic functions improves but the cost of quantum simulation for evaluation also increases. The number of top candidates chosen in each case is listed along the horizontal axis in the plots. Note that a reasonable fraction of cases produce a scaled probability less than one, indicating you would be better off just using the original circuit.
For grid instances, random (b) is the worst. Energy approximation (c) performs better than neural network (d) and is comparable (though still inferior to) to simulation (a). However, for complete graph, none of the heuristic functions is comparable to simulation (e).


\section{The Effect of Noise}

The purpose of this section is to analyze the effects of a simple noise model on the Gibbs objective function of Eq.~\eqref{eq-gibbs}.
In the absence of noise, the ideal Gibbs objective function is given by
$$
f_\text{ideal} = - \log \left\langle e^{-\eta E}\right\rangle_\psi,
$$
where $E$ is the Hamiltonian we are optimizing and the angled brackets represent the quantum expectation value in the output state $\psi$ of the quantum circuit ansatz.

The noise model we consider is a simple depolarizing channel. With probability $1-p$ the quantum circuit is executed perfectly and the output state is the one we expect. With probability $p$, there is some error in the execution and the output state is the maximally mixed one. In other words, with probability $p$ we sample from the uniform distribution on bit strings instead of the desired Born distribution. In the language of density operators, we can say that the effective density operator describing the quantum state is
$$
(1-p)|\psi\rangle\!\langle \psi | + \frac{p}{2^n}I,
$$
where $n$ is the number of qubits. 

Using this error model, we can ask what the noise does to the Gibbs objective function $f$.
We simply replace the expectation value in the ideal state $\psi$ with an expectation values in the noisy state $(1-p)|\psi\rangle\!\langle \psi | + pI/2^n$. Equivalently, we can take a weighted average of the $\langle \cdot \rangle_\psi$ expectation value with an expectation value according the uniform distribution over bit strings. We find
\begin{align*}
f_\text{noisy} &= - \log \left((1-p)\left\langle e^{-\eta E}\right\rangle_\psi+\frac{p}{2^n}\text{Tr}~e^{-\eta E}\right)\\
&= f_\text{ideal} - \log\left(1-p\frac{\left\langle e^{-\eta E}\right\rangle_\psi-\text{Tr}~e^{-\eta E}/2^n}{\left\langle e^{-\eta E}\right\rangle_\psi}\right)
\end{align*}

Note that one expects $\left\langle e^{-\eta E}\right\rangle_\psi\geq\text{Tr}~e^{-\eta E}/2^n$ if the circuit is properly trained, and so the correction makes the objective function larger (worse), as it should. We also have the following bound on the change in the objective function, coming from the positivity of $e^{-\eta E}$:
\begin{equation}\label{eq:fnoise}
f_\text{noisy}  - f_\text{ideal} \leq - \log(1-p).
\end{equation}
For small $p$ the right-hand-side is just $\approx p$. It's reasonable to expect that $\left\langle e^{-\eta E}\right\rangle_\psi\gg\text{Tr}~e^{-\eta E}/2^n$---in other words, the trained ansatz should have a much better Gibbs objective function value than the uniform distribution over bit strings---which means that the bound in Eq.~\eqref{eq:fnoise} will be approximately saturated.

This means that we can directly translate improvements to the objective function into resilience against depolarizing noise. An improvement of size $\Delta f$ in the objective function can counteract the effect of depolarizing noise with size $p \approx \Delta f$ (assuming $p\ll 1$).

It is also worth noting the effects of noise on the probability of finding a low-energy bit string, $P(E < E_0)$. Using the same depolarizing noise model,
$$
P_\text{noisy} = P_\text{ideal} - p(P_\text{ideal} - P_\text{uni}).
$$
Here $P_\text{uni}$ is just the probability for success by random guessing using the uniform distribution on bit strings. Then, taking logs, we find
$$
\log P_\text{noisy} = \log P_\text{ideal} + \log\left(1-p\frac{P_\text{ideal}-P_\text{uni}}{P_\text{ideal}}\right)
$$
This is very similar to what we saw in for the behavior of the Gibbs objective function. This is not a coincidence: part of the reason why the Gibbs objective function was chosen is that the operator $e^{-\eta E}$ for appropriate values of $\eta$ behaves very similarly to the projection operator one would use to define $P$. For large $\eta$ and $E_0$ close to $E_\text{gs}$, $P$ and $\langle e^{-\eta E} \rangle_\psi$ become equal up to a state-independent multiplicative factor.

\section{Relative Improvement of the Probability of Low Energy and Reduction of the Number of Two-qubit Gates}

In Table~\ref{table:relative_improvement}, we report the relative improvement of the probability of low energy and reduction of the number of two-qubit gates compared to the usual prescription of the QAOA (QAOA$+$energy) for 1000 grid instances and 1000 complete graph instances.
The relative improvement of the probability of low energy is
\begin{equation}\label{eq:relative_proba}
\left(\frac{P_{\{\mathrm{ansatz}\}+\mathrm{Gibbs}}(E < 0.95E_\text{gs})}{P_{\mathrm{QAOA}+\mathrm{energy}}(E < 0.95E_\text{gs})}-1\right) \times 100\%.
\end{equation}
For sparse$+$Gibbs, the ansatz for each instance is the ansatz with the lowest Gibbs objective function value in AAS by removing up to 20 two-qubit gates. The relative reduction of the number of two-qubit gates is
\begin{equation}\label{eq:relative_two_qubit_gates}
\left(\frac{N(\mathcal{A}_{\{\mathrm{ansatz}\}+\mathrm{Gibbs}})}{N(\mathcal{I})} - 1\right) \times 100\%,
\end{equation}
where $N(\cdot)$ counts the number of edges in the ansatz graph. For the usual prescription of the QAOA, $\mathcal{A}_{\{\mathrm{ansatz}\}+\mathrm{energy}}=\mathcal{I}$, so the relative reduction is always $0\%$.
QAOA$+$Gibbs brings $10.8\%$ and $8.6\%$ median relative improvement of the probability of low energy for grid and complete graph instances, respectively. By using a sparse ansatz together with the Gibbs objective function, the median relative improvement of the probability of low energy is $44.4\%$ and $244.7\%$, with reduction of the number of two-qubit gates by $20.8\%$ and $33.3\%$, for grid and complete graph instances, respectively.

\begin{table*}[htb]
\centering
\begin{tabular}{c  c  ccc c ccc } 
\toprule[0.1em]
\multirow{2}{*}{Instance Type} & \multirow{2}{*}{Prescription}  & \multicolumn{3}{c}{Relative Probability of Low Energy (Eq.~\ref{eq:relative_proba})} & & \multicolumn{3}{c}{Relative Number of Two-Qubit Gates (Eq.~\ref{eq:relative_two_qubit_gates})} \\
\cmidrule[0.05em](lr){3-5}\cmidrule[0.05em](lr){7-9}
 & & 5th percentile & median & 95th percentile & & 5th percentile & median & 95th percentile  \\ 
\midrule[0.05em]
\multirow{2}{*}{Grid} & QAOA$+$Gibbs & $+5.9\%$ & $+10.8\%$ & $+17.5\%$ & & $0\%$ & $0\%$ & $0\%$  \\ 
                      & sparse$+$Gibbs & $+15.7\%$ & $+44.4\%$ & $+102.7\%$ & & $-54.2\%$ & $-20.8\%$ & $-8.3\%$  \\
\midrule[0.05em]
\multirow{2}{*}{Complete} & QAOA$+$Gibbs & $+3.4\%$ & $+8.6\%$ & $+18.7\%$ & & $0\%$ & $0\%$ & $0\%$  \\ 
                      & sparse$+$Gibbs & $+114.4\%$ & $+244.7\%$ & $+485.6\%$ & & $-44.4\%$ & $-33.3\%$ & $-24.4\%$  \\ 
\bottomrule[0.1em]
\end{tabular}
\caption{Relative improvement of the probability of low energy (Eq.~\ref{eq:relative_proba}) and reduction of the number of two-qubit gates (Eq.~\ref{eq:relative_two_qubit_gates}) compared to the usual prescription of the QAOA for 1000 grid instances and 1000 complete graph instances. The values of 5th percentile, median and 95th percentile for reported for two instance types.}\label{table:relative_improvement}
\end{table*}

In Figure~\ref{fig:plot_more_instances_grid} and \ref{fig:plot_more_instances_complete}, we randomly sample five instances out of 1000 for each instance type and show the structure of their associated QAOA and best sparse ansatzes with the Gibbs objective function.

\begin{figure}[htb]
\includegraphics[width=\columnwidth]{fig/plot_grid_4x4_instances_0_no_couplings_text}
(a)
\includegraphics[width=\columnwidth]{fig/plot_grid_4x4_instances_100_no_couplings_text}
(b)
\includegraphics[width=\columnwidth]{fig/plot_grid_4x4_instances_200_no_couplings_text}
(c)
\includegraphics[width=\columnwidth]{fig/plot_grid_4x4_instances_300_no_couplings_text}
(d)
\includegraphics[width=\columnwidth]{fig/plot_grid_4x4_instances_400_no_couplings_text}
(e)
\caption{\label{fig:plot_more_instances_grid} Five instances of random couplings and the structures of the associated QAOA and best sparse ansatzes for grid problems with the Gibbs objective function.
On the left, each edge in the instance graph is colored by its coupling from blue ($-1$) to red ($1$).
We show the relative improvement of the probability of low energy and reduction of the number of two-qubit gates compared to the usual prescription of the QAOA.}
\end{figure}

\begin{figure}[htb]
\includegraphics[width=\columnwidth]{fig/plot_complete_10_instances_0}
(a)
\includegraphics[width=\columnwidth]{fig/plot_complete_10_instances_100}
(b)
\includegraphics[width=\columnwidth]{fig/plot_complete_10_instances_200}
(c)
\includegraphics[width=\columnwidth]{fig/plot_complete_10_instances_300}
(d)
\includegraphics[width=\columnwidth]{fig/plot_complete_10_instances_400}
(e)
\caption{\label{fig:plot_more_instances_complete} Five instances of random couplings and the structures of the associated QAOA and best sparse ansatzes for complete graph problems with the Gibbs objective function.
On the left, each edge in the instance graph is colored by its coupling from blue ($-1$) to red ($1$).
We show the relative improvement of the probability of low energy and reduction of the number of two-qubit gates compared to the usual prescription of the QAOA.}
\end{figure}

\bibliography{references}

\end{document}